\documentclass[]{aa}  

\usepackage{graphicx}
\usepackage{txfonts}
\usepackage{natbib}
\bibpunct{(}{)}{;}{a}{}{,} % to follow the A&A style
\def\ltsim{\raise 2pt \hbox {$<$} \kern-0.9em \lower 2pt \hbox {$\sim$}}
\def\gtsim{\raise 2pt \hbox {$>$} \kern-1.1em \lower 4pt \hbox {$\sim$}}

\begin{document} 

\title{Radio-continuum surveys with SKA and LOFAR: \\
a first look at the perspectives for radio mini-halos}

\author{M. Gitti\inst{1,2}, G. Brunetti \inst{2}, R. Cassano\inst{2},
  S. Ettori\inst{3,4}}

\institute{ 
Dipartimento di Fisica e Astronomia (DIFA), Universit{\`a} di Bologna, via Gobetti 93/2, 40129 Bologna, Italy \\ \email{myriam.gitti@unibo.it}
\and
Istituto Nazionale di Astrofisica (INAF) -- Istituto di
Radioastronomia (IRA), via Gobetti 101, I-40129
Bologna, Italy 
\and
Istituto Nazionale di Astrofisica (INAF) -- Osservatorio di Astrofisica e Scienza dello Spazio (OAS), via Gobetti 93/3,
I-40129 Bologna, Italy
\and
Istituto Nazionale di Fisica Nucleare (INFN) -- Sezione di Bologna, viale Berti Pichat 6/2, I-40127 Bologna, Italy}

\authorrunning{Gitti et al.} 
\titlerunning{Radio mini-halos in future radio-continuum surveys}

\date{Accepted }

\abstract 
{%Context:
  Diffuse synchrotron radio emission has been observed in a number of
  cool-core clusters on scales comparable to that of the cooling
  region. These radio sources are called `mini-halos'. In order to
  understand their origin, which is still unclear, joint radio and
  X-ray
%  \LEt{the slash commonly means "ratio" as in "S/N" for
%  "signal-to-noise ratio". You appear to be using it like the in the
%  Results paragraph of this abstract, but here it seems to me you
%  mean "and". If this is not the case, please change this back here,
%  but please check throughout that you use the slash to mean "ratio"
%  and not "and" or "or" to avoid confusion for your readers}
  statistical studies of large cluster samples are necessary to
  investigate the radio mini-halo properties and their connection with
  the cluster thermodynamics. }
{%Aims:
  We here extend our previous explorative study and investigate the
  perspectives offered by surveys in the radio continuum with the LOw
  Frequency ARray (LOFAR) and the Square Kilometre Array (SKA), in
  particular examining the effect of the intracluster magnetic field
  in the mini-halo region for the first time. }
{%Methods:
  By considering the minimum flux detectable in radio surveys and
  exploiting the $P_{\rm radio}$-$L_{\rm X}$ correlation observed for
  known mini-halos, we estimate the detection limits achievable by
  future radio observational follow-up of X-ray cluster samples, such
  as HIFLUGCS and eROSITA.  This allows us to estimate the maximum
  number of radio mini-halos that can potentially be discovered in
  future surveys as a function of redshift and magnetic field
  strength.}
{%Results:
  Under the optimistic assumption that all cool-core systems host a
  mini-halo and that the radio vs. X-ray scaling relation extends to
  systems with lower X-ray luminosity, we show that future radio
  surveys with LOFAR and SKA1 (at $\sim$140 MHz and $\sim$1.4~GHz)
  have the potential to discover $\sim$1,000-10,000 radio mini-halo
  candidates up to redshift $z=1 $. This shows that these surveys may
  be able to produce a breakthrough in the study of these sources.  We
  further note that future SKA1 radio surveys at redshift $z>0.6$ will
  allow us to distinguish between different magnetic fields in the
  mini-halo region, because higher magnetic fields are expected to
  produce more powerful mini-halos, thus implying a larger number of
  mini-halo candidates detected at high redshift.  For example, the
  non-detection with SKA1 of mini-halos at $z>0.6$ will suggest a low
  magnetic field ($B <$ few $\mu$G).  The synergy of these radio
  surveys with future X-ray observations and theoretical studies is
  essential in establishing the radio mini-halo physical nature.}  {}


\maketitle

\section{Introduction} 

Radio mini-halos are diffuse, faint (at a level of few $\mu$Jy
arcsec$^{-2}$ at 1.4~GHz), amorphous (almost roundish) radio sources
with a steep ($\alpha>1$, where the flux density at the frequency
$\nu$ is $S_{\nu} \propto \nu^{-\alpha}$)
  % \LEt{it is unusual to have discursive footnotes in a paper (urls
  % and such are okay). Please move all your footnotes except for
  % footnote 5 to the main text}.
  radio spectral index \citep[e.g.,][for an observational
  review]{Feretti_2012}. They are extended on scales (total size)
  $\sim$100-500 kpc at the center of a number of cool-core (CC)
  clusters, tracing regions where the cooling time of the intracluster
  medium (ICM) is short.  The mini-halo diffuse emission is always
  observed to surround the intense radio emission of the brightest
  cluster galaxy (BCG), which often shows nonthermal radio jets and
  lobes ejected by the central active galactic nucleus (AGN). The lobe
  radio plasma interacts strongly with the ICM, which is clearly
  spatially separated from it, by inflating large X-ray cavities and
  triggering the so-called `radio-mode AGN feedback' \citep[e.g.,][for
  a review]{Gitti_2012}.  Although the central radio-loud AGN is
  likely to play a role in the initial injection of the relativistic
  particles emitting in the mini-halo region, the diffuse radio
  emission is thought to be not directly connected with the BCG radio
  bubbles, but is believed to be truly generated from the ICM on
  larger scales, where the thermal and nonthermal components are mixed
  \citep[e.g.,][for a review]{Brunetti-Jones_2014}.

Similarly to the well-known case of giant halos extended on
cluster-scale, the problem of the origin of radio mini-halos stems
from the fact that the time necessary for the emitting electrons to
diffuse efficiently on the scales covered by the radio emission is
longer than their radiative lifetime.
To overcome this problem, {\it \textup{in situ}} particle reacceleration by
turbulence in the CC region \citep[][]{Gitti_2002, Gitti_2004,
  Mazzotta-Giacintucci_2008, ZuHone_2013}, or alternatively, hadronic
models in which secondary electrons are continuously generated by
$p$-$p$ collisions in the cluster volume
\citep[][]{Pfrommer-Ensslin_2004, Jacob-Pfrommer_2017b}, have been
invoked \citep[][for a review]{Brunetti-Jones_2014}.

Nevertheless, the origin of mini-halos is still debated since several
fundamental questions remain unresolved. For instance, the possible
connection between the origin of mini-halos and the heating of the gas
in the central region of CC clusters is still an open issue.
Turbulence in CCs may be generated by several mechanisms, including
the interplay between the outflowing relativistic plasma in AGN jets
and lobes, and sloshing gas motions \citep[][]{Fujita_2004,
  Heinz_2006, Mazzotta-Giacintucci_2008, ZuHone_2013,
  Zhuravleva_2016}.
Interestingly, turbulence has been measured in the Perseus cluster
with properties (strength and scales) that appear compatible with
those necessary to explain the origin of the mini-halo with reacceleration
models \citep{Hitomi_2016, Hitomi_2017}.  At the same time,
turbulent heating has been suggested to play an important role in the
thermal balance of galaxy cluster plasmas
\citep{Dennis-Chandran_2005, Gaspari_2012, Zhuravleva_2014}.  This
raises the fascinating possibility of a common origin of radio
mini-halos and gas heating in CC, provided that 
the same turbulence is channelled partly into gas heating and partly
into reacceleration of relativistic particles \citep{Bravi_2016}.
Conversely, the same problem of the origin of mini-halos and the
heating of gas in CCs can be addressed in hadronic models through the
dissipation of cosmic-ray induced instabilities into the ICM
\citep{Fujita-Ohira_2012, Jacob-Pfrommer_2017b}.
In contrast to the case of giant radio halos, the gamma-rays induced by
the decay of neutral pions that are generated in hadronic models for
mini-halos is still compatible with current gamma-ray limits derived
for nearby CCs \citep[e.g.,][]{Ahnen_2016, Jacob-Pfrommer_2017b}.

Another open problem is generally the connection between mini-halos
and gas dynamics of the CC and of the hosting clusters. Signatures of
minor dynamical activity have been detected in a number of clusters
hosting mini-halos \citep{Gitti_2007b, Cassano_2008b}, and a
morphological connection between mini-halos and cold fronts has been
shown in several cases \citep[e.g.,][]{Mazzotta-Giacintucci_2008,
  Giacintucci_2014b, Marielou_2017}. Furthermore, large-scale radio emission in the
form of giant radio halos is also detected in a few CC clusters
showing minor dynamical activity \citep[e.g.,][]{Sommer_2016},
suggesting a possible connection or evolution between giant halos and
mini-halos \citep[][]{Zandanel_2014, Brunetti-Jones_2014}.

Unfortunately, current observations of mini-halos do not allow us to
address these open questions.  Obviously, one problem is the limited
statistics that prevent us from drawing firm conclusions on the
connection between thermal and nonthermal phenomena in CCs.  For this
reason, \citet{Gitti_2015} used scaling relations between the
properties of mini-halos and of the hosting CC clusters to start
investigating the capabilities of the Square Kilometre Array
\citep[SKA,][]{Braun_2015} in the detection of mini-halos at mid
frequencies ($\sim 1$ GHz).

In this paper, we extend the previous work to low radio frequencies
and derive expectations assuming different magnetic fields in the
mini-halo region.  We focus in particular on the case of the LOw
Frequency ARray \citep[LOFAR,][]{lofar_2013} and SKA-LOW at
$\sim 140 $ MHz, and also discuss the synergy with current and future
X-ray missions such as eROSITA \citep{erosita_2012}.  The questions we
would like to address are essentially whether these next-generation
radio telescopes may be expected to significantly increase the statistics
of these radio sources, detecting mini-halos in less X-ray luminous
systems and at higher redshift, and to constrain the intracluster
magnetic field in the mini-halo region.

We adopt a $\mathrm{\Lambda CDM}$ cosmology with $\mathrm{H_{0}=70}$
km $\mathrm{s^{-1}Mpc^{-1}}$, $\Omega_{M} = 1 - \Omega_{\Lambda} =
0.3$.

\section{Mini-halo sample: current statistics}
\label{statistics.sec}

The detection of radio mini-halos is complicated by the need to
separate their faint, low surface brightness emission from the bright
emission of the radio-loud BCG embedded in it.  This requires
observations performed with a technical setup ensuring high
sensitivity to diffuse emission, provided by short baselines, but at
the same time allowing an accurate subtraction of discrete sources,
which are best identified by long baselines.  Therefore, good spatial
dynamic ranges, along with dynamic ranges much higher ($\gg 1000$)
than those in available surveys, are typically required to properly
image the radio mini-halo emission.
For this reason, the mini-halo detection has indeed proven to be very
challenging with the current facilities, even with pointed
observations \citep[e.g.,][]{Govoni_2009, Giacintucci_2014a}.  A first
statistical assessment of the occurrence of mini-halos in a
mass-selected ($M_{500}>6\times 10^{14} M_{\odot}$) sample of galaxy
clusters found that all mini-halos are hosted by CC clusters, defined
as clusters with a value of the central entropy $K_0 <30$ keV cm$^2$
(where $K_0 = kT_0 n_0^{-2/3}$, $n_0$ and $T_0$ being the central
values of the ICM electron density and temperature, respectively), and
that almost all CC clusters of the sample, namely $\sim$80\%, host a
mini-halo \citep{Giacintucci_2017}.  This fraction is similar to the
fraction of CC clusters whose BCG hosts a radio AGN
\citep[70\%,][]{Burns_1990,Dunn-Fabian_2006,Best_2007,Mittal_2009}, in
line with the observed evidence that AGN are ubiquitous in mini-halos
\citep[e.g.,][]{Giacintucci_2014a}.

\begin{figure}
%\hspace{0.1in}
\centering
\includegraphics[bb=18 144 592 718, width=0.46\textwidth]{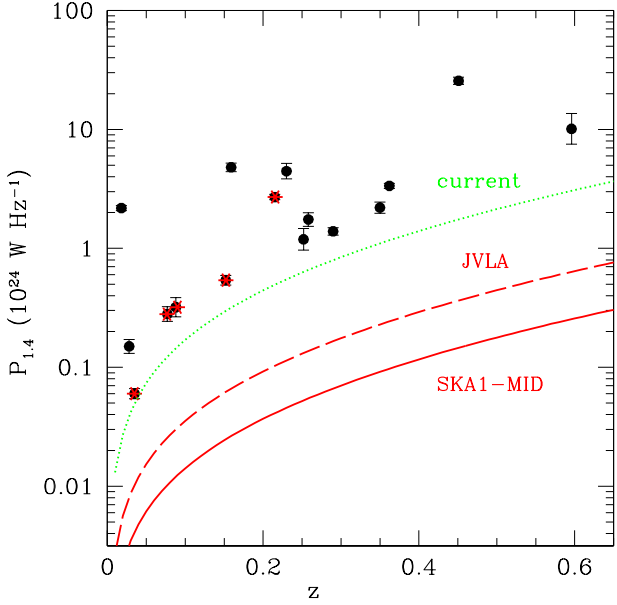}
%\vspace{0.5cm}
\caption{\label{Pradio.fig} Radio power at 1.4~GHz
  \citep[from][]{Giacintucci_2014a, vanWeeren_2014} vs. redshift of
  the sample of 16 confirmed mini-halos known up to 2016 (sorted by decreasing radio power at 1.4~GHz):
RX~J1347.5$-$1145 (z=0.451), 
Phoenix (z=0.596), 
RX~J1720.1$+$2638 (z=0.159),
A~2390 (z=0.23), 
RX~J1532.9$+$3021 (z=0.362), 
RXC~J1504.1$-$0248 (z=0.215),
RBS~797 (z=0.35), 
Perseus (z=0.018),
MS~1455.0$+$2232 (z=0.258), 
ZwCl~3146 (z=0.290),
A~1835 (z=0.252), 
A~2204 (z=0.152), 
A~478 (z=0.088), 
A~2029 (z=0.077), 
Ophiuchus (z=0.028), 
and 2A~0335$+$096 (z=0.035).
The clusters belonging
  to the HIFLUGCS sample are highlighted with red stars.  The green
  dotted line is indicative of the existing detection limit reached in
  the literature with pointed observations.  The red dashed and solid
  lines are the detection limits achievable with JVLA and SKA1-MID,
  respectively, with the performances adopted in Table 1 (see
  Sect. 3.1).  }
%\vspace{1in}
\end{figure}

\begin{figure}[t]
\centering 
\includegraphics[bb=18 144 592 718, width=.46\textwidth,angle=0]{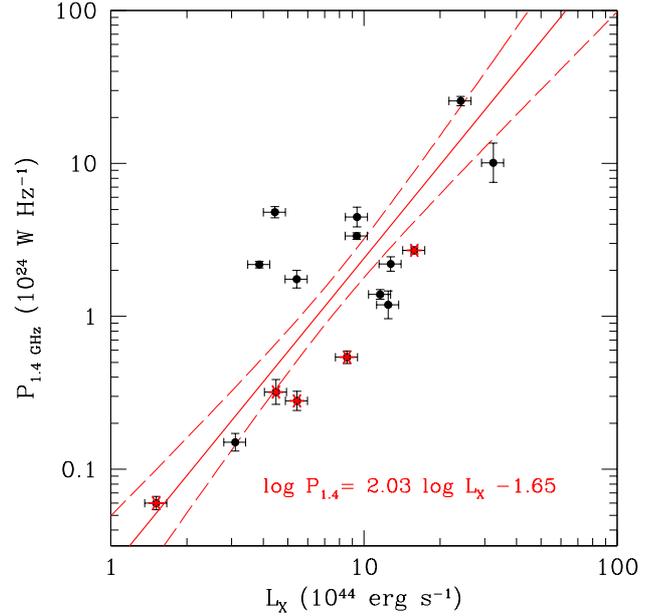}
%\vspace{-0.1cm}
\caption{\label{correlation.fig} Radio power at 1.4~GHz versus the
  X-ray luminosity inside $R_{500}$ \citep[taken from][and converted into
    the 0.5-2.0 keV band]{Piffaretti_2011} for the mini-halo sample. The red solid
  line is the best-fit relation (see Eq. \ref{Pradio-Lx.eq}), enclosed
  by the 1$\sigma$ confidence lines (red dashed). The clusters
  belonging to the HIFLUGCS sample are highlighted with red stars. }
\end{figure}

In Figure \ref{Pradio.fig} we show the $K$-corrected radio power
  at 1.4~GHz versus redshift distribution of the sample of {\it
  \textup{confirmed}} mini-halos known up to 2016, which comprises 16 objects
selected from the mini-halo collection reported in \citet[][and
references therein]{Giacintucci_2014a} by excluding the objects that
they classify as `candidate' or `uncertain' and including the new
detection in the Phoenix cluster \citep{vanWeeren_2014}.
We quantitatively checked that the 16 mini-halos in Figure 1 are
hosted in CC clusters; by taking values from the literature
\citep{Cavagnolo_2011, Hudson_2010, McDonald_2012}, we find that all
mini-halo clusters have central entropy $K_0$ \ltsim 25 keV cm$^2$.
According to \cite{Hudson_2010}, this definition applies to strong
cool-core (SCC) clusters.
The green dotted line in Figure 1 is indicative of the current
mini-halo detection limit obtained by assuming angular resolution
$\theta_b = 5$ arcsec and sensitivity = 30 $\mu$Jy beam$^{-1}$, typical of
pointed observations with the VLA in B-array (see Sect. 3.1 for
details on the criterion adopted to derive the minimum detectable
flux).  It is evident that our current ability of detecting mini-halos
is limited and that we may be missing many faint mini-halos.

One possibility for estimating the perspective offered by the SKA and its
pathfinders in the ability of detecting radio mini-halos is to
first link the nonthermal properties of these sources to the thermal
properties of their host galaxy clusters.  
For the above sample of 16 confirmed mini-halos, we considered the
X-ray luminosity taken from the MCXC catalog of
\citet{Piffaretti_2011} \citep[for Phoenix, which is the only cluster
not included in the MCXC, the luminosity was taken
from][]{McDonald_2012}.  This catalog is based on publicly available
ROSAT All Sky Survey data, which have then been systematically
homogenized to an overdensity of 500.  We converted the tabulated
X-ray luminosity from 0.1-2.4 keV into 0.5-2 keV band assuming an
Xspec {\ttfamily apec} plasma model at the observed cluster
temperature, redshift, and metallicity \citep[taken
from][]{Cavagnolo_2009, Hudson_2010}.  In these energy bands, the
signal is almost independent of any temperature and is only
proportional to the square of the gas density. Therefore, the
conversion itself introduces a negligible systematic error in the
total error budget of about 10\% that we propagate through our
statistical analysis.  The resulting average conversion factor for our
sample is $L_{0.5-2.0} \sim 0.6 \, L_{0.1-2.4}$.  The choice of this
particular energy band was made for consistency with the method used
in Section 3.3 to estimate the radio luminosity function of mini-halos
by linking it to the X-ray luminosity function of galaxy clusters,
which is measured in the 0.5-2.0 keV energy band \citep{Mullis_2004}.

We found a correlation between the 1.4~GHz radio power of the
mini-halos, $P_{1.4}$ (in units of $10^{24}$ W Hz$^{-1}$), and the
global X-ray luminosity of the host clusters in the 0.5-2.0 keV band,
$L_{\rm X} $ (in units of $10^{44}$ erg s$^{-1}$), in the form
\begin{equation}
\log P_{1.4} = a \, \log L_{\rm X} + b
\label{Pradio-Lx.eq}
,\end{equation}
where $a = 2.03 \pm 0.20$ and $ b = - 1.65 \pm 0.21 $ are the best-fit
values derived by using the bivariate correlated error and intrinsic
scatter (BCES) algorithm \citep[]{Akritas_1996} to perform regression
fits (bisector method). The best-fit correlation is shown in Figure
\ref{correlation.fig}, along with the lines enclosing the area that
has a 1$\sigma$ chance of containing the true regression line
\citep[estimated from Eq. 7 in][]{Cassano_2013}.

To ensure that the trend observed in the radio luminosity -- X-ray
luminosity plane is not biased by any effect due to the underlying
dependence of these two properties on redshift, we checked that a
correlation is present in the radio flux -- X-ray flux plane as well. We
found a Spearman rank correlation coefficient of $r_s \sim$ 0.6 in
both planes, with a probability of no correlation of $probs \sim$ 1\%,
thus indicating that the relation between mini-halo radio power and
X-ray luminosity may be intrinsic.

\begin{figure}[t]
\centering 
\includegraphics[bb=18 144 592 718,width=.46\textwidth,angle=0]{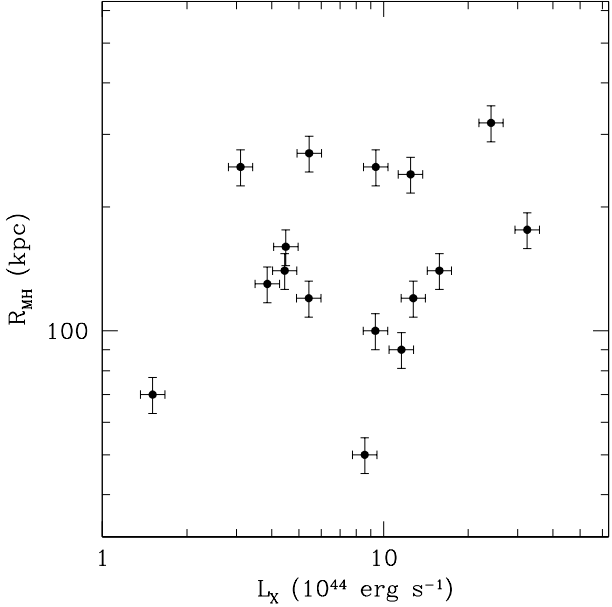}
%\vspace{-0.1cm}
\caption{\label{rmh.fig} Average radius of the mini-halo
    \citep[from][]{Giacintucci_2014a, Bravi_2016} vs. the X-ray
    luminosity inside $R_{500}$ \citep[taken from][and converted into
    the 0.5-2.0 keV band]{Piffaretti_2011}. No obvious correlation is
  visible (we found $r_s \sim$0.2, with a probability of no
  correlation of $probs \sim$40\%). }
\end{figure}

\section{Role of future radio surveys}

\subsection{Detecting radio mini-halos in galaxy clusters}
\label{detecting.sec}

In order to evaluate the role of future radio surveys in the ability
to discover new radio mini-halos, we need to estimate the minimun flux
of diffuse radio emission that can be detected with given
observational performances. To do so, we assumed an exponential
distribution of the diffuse radio brightness \citep{Murgia_2009} and
determined the region that contains a significant fraction of the
integrated flux density.  In particular, assuming a radial profile in
the form $I(r) = I_0 \, e^{-r/r_e}$, where $I_0$ is the central radio
surface brightness and $r_e$ is the $e$-folding radius (i.e., the
radius at which the brightness drops to $I_0/e$), radio mini-halos
emit about half of their total radio flux within a radius
$r_{\rm half} \sim 1.68 \, r_e$, which is called the half-radius.  In
order to estimate the minimum flux that can be detected with
observations performed at given resolution and sensitivity levels, we
adapted to mini-halos the criterion of \citet{Cassano_2012} based on a
threshold in flux for giant halos, and assumed that mini-halos can be
detected when the integrated flux within their half-radius gives a
signal-to-noise ratio $\xi_2$.  In particular, the minimum detectable
flux as a function of redshift, $f_{\rm min}(z)$, is
\begin{equation}
f_{\rm min}(z)\simeq2.86\times10^{-3}\, \xi_2\,
\Big(\frac{\mathrm{F_{\rm rms}}}{10 \, \mu\mathrm{Jy}}\Big)
\Big(\frac{10''}{\theta_{\rm b}}\Big)
\Big(\frac{\theta_{\rm half}(z)}{1''}\Big) \; \; \; \; [{\rm mJy}]
\label{fmin.eq}
,\end{equation}
\noindent
where ${\rm F}_{\rm rms}$ and $\theta_{\rm b}$ are the rms noise per
beam and the beam angular size (in arcsec) of the observations,
respectively, and $\theta_{\rm half}(z)$ is the angular size (in
arcsec) of mini-halo half radius at a given redshift. 

We aim at evaluating the possible detections of mini-halos with a
typical size of $\sim$200 kpc \citep[e.g.,][]{Giacintucci_2014a},
therefore we assume a representative half-radius of $\sim$100 kpc.
This is in line with the lack of a clear correlation between the
mini-halo size and other cluster properties, such as the X-ray
luminosity, as shown in Figure \ref{rmh.fig}.  We did not find any
obvious correlation between the mini-halo radius and the 1.4~GHz radio
power either ($r_s \sim$0.2, with a probability of no correlation of
$probs \sim$40\%), which is in agreement with the observed $P_{1.4}$ -
$L_{X}$ correlation.  Therefore, $\theta_{\rm half}$ is a function of
redshift alone.
As an example, at redshift $z=0.6,$ the angular half-radius of a
typical radio mini-halo is $\theta_{\rm half} \sim$15
arcsec. Following \citet{Cassano-SKA}, we further adopt $\xi_2 =7$ as a
reference value.
 
We here consider the radio observational performances reported
in Table 1: at mid frequency, the SKA1-MID all sky survey at
$\sim$1.4~GHz \citep{Prandoni-SKA}, along with JVLA pointed
observations in L band (1-2 GHz), and at low frequency, the LOFAR
Two-metre Sky Survey \citep[LoTSS,][]{Shimwell_2017} and SKA1-LOW all-sky surveys at $\sim$140~MHz \citep{Prandoni-SKA}.  In particular, we
estimated the minimum flux as described above by assuming values of
SKA1 surveys at confusion limit \citep[see][for a more detailed
discussion]{Cassano-SKA}.
We did not consider the upcoming VLA Sky Survey
(VLASS)\footnote{{\it
    https://science.nrao.edu/enews/8.11/index.shtml\#vlass}} at
2.5-arcsec resolution because owing to the relatively high frequency
(S-band: 2-4 GHz) and low sensitivity (120 or 70 $\mu$Jy beam$^{-1}$
achievable with 3 passes or 1 pass, respectively), it is not expected
to be competitive in detecting mini-halos.

The value of $\theta_b$ in Table 1 can be obtained after tapering from
higher-resolution images. For example, the angular
resolution of a few arcseconds that can be achieved at medium frequency can be used to produce
BCG-subtracted images, which can be then tapered up to a lower
resolution to increase the sensitivity to the extended emission.

\begin{table}
\begin{center}
  \caption{ Observational performances adopted in this work.  }
\begin{tabular}{lcccc}
\hline
\hline
Instrument & mode & $\nu$ & ${\rm F}_{\rm rms}$ & $\theta_b$\\
~ & ~ & (MHz) & ($\mu$Jy beam$^{-1}$) & ($''$) \\
\hline
\hline
JVLA  & pointed & 1400 & 10 & 8 \\
SKA1-MID  & survey ($3 \pi$) & 1400 & 4 & 8 \\
\hline
LOFAR & survey ($3 \pi$) & 140 & 200 & 8 \\
SKA1-LOW & survey ($3 \pi$) & 140 & 20 & 8 \\
\hline
\hline
\label{survey.tab}
\end{tabular}
\vspace{-0.15in}
\begin{minipage}{0.45\textwidth}
    \footnotesize \textbf{Notes:} 
  ${\rm F}_{\rm rms}$ is the sensitivity, $\theta_b$ is
  the tapered resolution. At $\nu \sim$1.4~GHz, the survey speed of SKA1-MID
  is more than two orders of magnitude faster than JVLA, whereas at
  $\nu \sim$140~MHz the survey speed of SKA1-LOW is more than one order of
  magnitude faster than LOFAR\footnote{{\it https://astronomers.skatelescope.org/documents}}.
\end{minipage}
\end{center}
\end{table}

From the minimum detectable flux (Eq. \ref{fmin.eq}), we estimated the
minimum detectable radio power at each redshift calculated as
$ P_{\nu}(z)=4 \pi \, D_{\rm L}^2 \, f_{\rm min, \nu}(z) \,
(1+z)^{(\alpha -1)} $, where $D_{\rm L}$ is the luminosity distance,
and $(1+z)^{(\alpha -1)}$ is the \textit{\textup{K-correction}} term, thus
obtaining a survey sensitivity in terms of radio power.
The results on the 1.4~GHz detection limit are shown in Figure
\ref{Pradio.fig}, where it is evident that the JVLA represents 
  already a great improvement with respect to existing
observations. We note, however, that such performances require pointed
observations. According to the JVLA exposure calculator, confirmed by
our direct experience on real observations
\citep[e.g.,][]{Ignesti_2017}, the values adopted in Table 1 are
achievable with approximately two-hour pointings. On the other hand,
SKA1-MID surveys will be able to detect all mini-halos with $P_{1.4}$
\gtsim ~$2\times 10^{23}$ W Hz$^{-1}$ up to redshift 0.6.

\subsection{Radio follow-up of X-ray cluster samples}
\label{followup.sec}

\begin{figure*}
\centerline{
\hspace{-0.7cm}
\includegraphics[width=.46 \textwidth,angle=0,origin=c,trim=1 1 1 1,
clip]{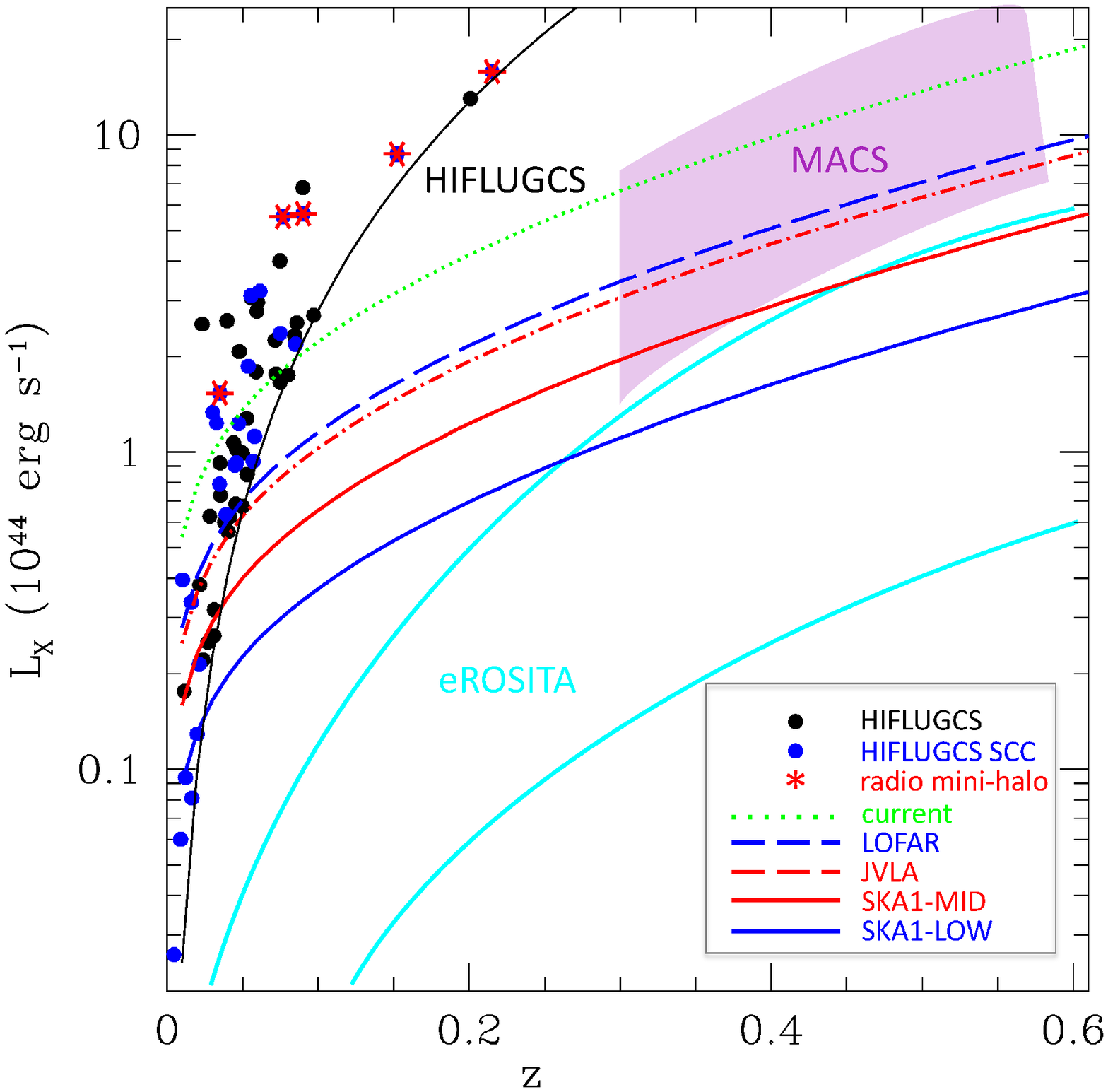}
\hspace{0.7cm}
\includegraphics[width=.46 \textwidth,angle=0,origin=c,trim=1 1 1 1, clip]{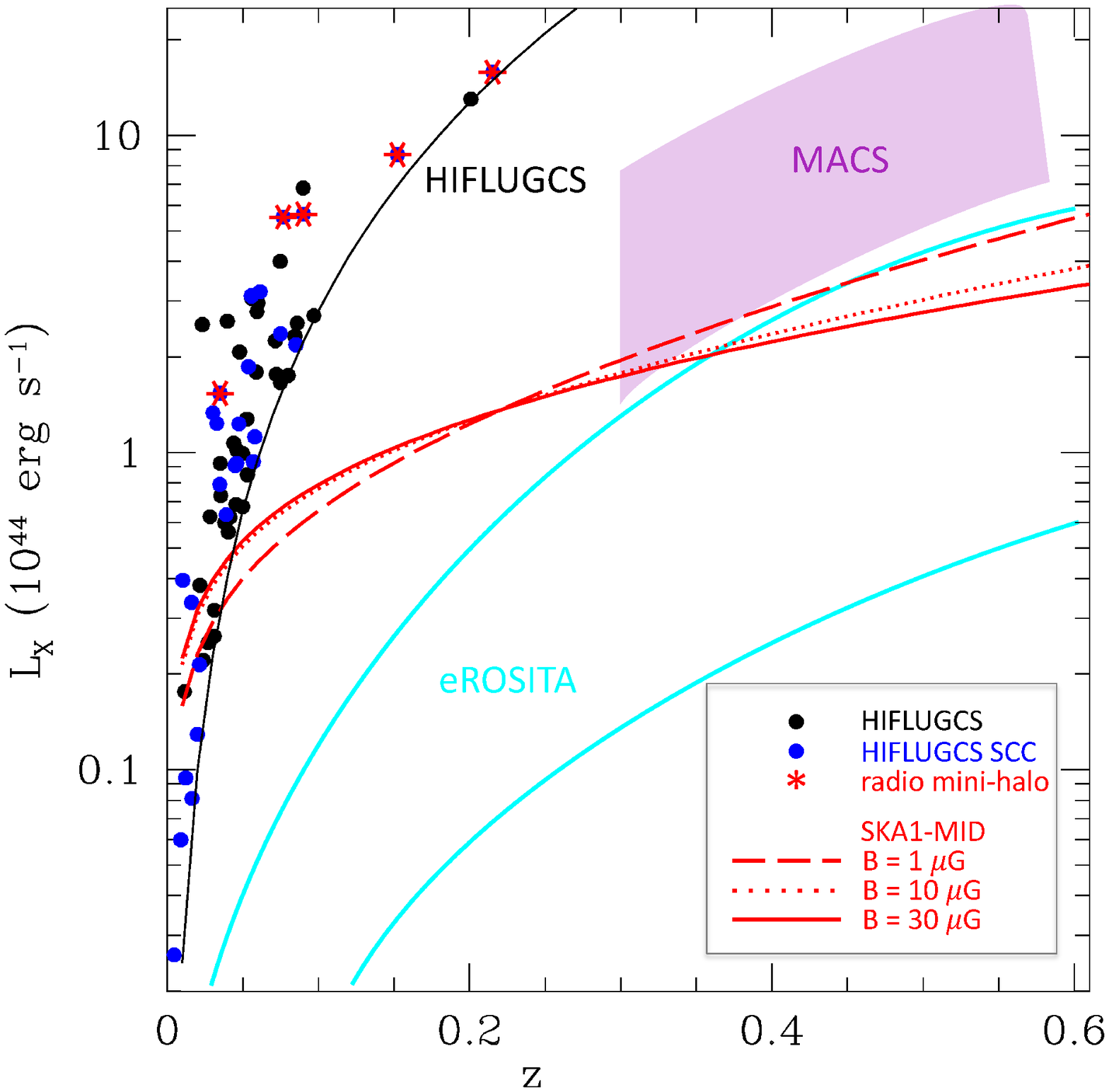}
}
\caption{\label{statistics.fig} {\it Left panel:} X-ray luminosity in
  the 0.5-2.0 keV band vs. redshift for the clusters in the
  HIFLUGCS sample \citep{hiflugcs_2002}. The flux limit is shown
    as a black solid line. The blue filled circles represent the
  clusters classified as SCC \citep{Hudson_2010}, and the clusters
  known to host radio mini-halos are highlighted with red stars.  For
  comparison, we also draw the $L_{\rm X}$-$z$ distribution of
  clusters from other X-ray selected samples: the shaded magenta
  region indicates the position of MACS clusters \citep{macs_2010},
  and the cyan solid lines show the eROSITA threshold luminosity
  predicted with 500 (upper line) and 50 (lower line) photon counts
  \citep[][]{erosita_2012}.  Our results on the detection limits
  reachable by radio follow-up are estimated with the performances
  adopted in Table 1 (see Sect. 3.1) and a typical magnetic field of
  $B=$1 $\mu$G: from existing observations (green dotted line), from
  1.4~GHz JVLA pointed observations and SKA1-MID surveys (red dashed
  and solid lines, respectively), and from 140~MHz LOFAR and SKA1-LOW
  surveys (blue dashed and solid lines, respectively).  {\it Right
    panel:} Same as left panel, but showing only the detection limits
  reachable by radio follow-up with SKA1-MID assuming three different
  magnetic fields: $B=1 \, \mu$G (dashed line), $B=10 \, \mu$G (dotted
  line), and $B=30 \, \mu$G (solid line).  }
\end{figure*}

We further aim at investigating the radio follow-up of X-ray cluster
samples by adopting the basic assumption that mini-halos follow the
observed $P_{1.4}$-$L_X$ correlation (Eq. \ref{Pradio-Lx.eq}).
This correlation indicates a connection between the energy reservoir in
CC clusters and that associated with the nonthermal components
powering radio mini-halos \citep[see also][]{Bravi_2016}.
The nonthermal radiation, $P_{\rm NT}$, that can be maintained in the
region of radio mini-halos for a timescale that is longer than the
radiative life-time of the relativistic electrons is 
\begin{eqnarray}
P_{\rm NT} = P_{\rm radio} + P_{\rm IC} = P_{\rm radio}\left[1+ \left(
  \frac{B_{\rm CMB}}{B} \right)^2 \right]
,\end{eqnarray}
where $P_{\rm radio}$ is the synchrotron radiation, $P_{\rm
    IC}$ is the inverse Compton radiation, $B_{\rm CMB} = 3.25 \,
(1+z)^2 \, \mu$G is the magnetic field equivalent to the inverse
Compton losses with CMB photons, and $B$ is the magnetic field
intensity in the mini-halo region.
We can thus express the observed synchrotron radiation as a function
of the intracluster magnetic field and redshift:
$ \, P_{\rm radio} = P_{\rm norm} \, [B^2 / (B^2 + B^2_{\rm CMB})] \,
$, where the normalization $P_{\rm norm}$ is a function of $B$ and is
obtained by imposing that at the mean redshift of our sample
($<z> = 0.22$), the synchrotron radiation observed at 1.4~GHz follows
Eq. \ref{Pradio-Lx.eq}.

By matching the observed normalization of the correlation, we obtained
the general correlation (where we make the dependencies explicit)
\begin{equation}
P_{\rm 1.4} (B,z) = 10^b \, \left( \frac{B^2 + 23.4}{B^2 + 10.6 \, (1+z)^4} \right)
\, L^{a}_{\rm X} 
\label{norm.eq}
,\end{equation}
\noindent
where $P_{1.4}$ is in units of $10^{24}$ W Hz$^{-1}$, $L_{\rm X}$ is
in units of $10^{44}$ erg s$^{-1}$, $B$ is in $\mu$G, and $a$ and $b$
are the parameters in Eq. \ref{Pradio-Lx.eq}.
\\
By combining this correlation with the minimum detectable radio power,
estimated in Sect. \ref{detecting.sec} with the performances in Table
1, we finally obtained the X-ray luminosity thresholds that allow us
to investigate the possible radio follow-up coverage of (existing and
future) X-ray cluster samples.

In particular, we consider the X-ray HIFLUGCS sample \citep[HIghest
X-ray FLUx Galaxy Cluster Sample,][]{hiflugcs_2002}, which is a
flux-limited
($f_{X} \geq 2 \times 10^{-11}$ erg s$^{-1}$ cm$^{-2}$ in the 0.1-2.4
keV ROSAT band)
statistically complete sample of 64 galaxy clusters with mean redshift
$<$$z$$>$=0.05, with the largest coverage (100\%) of high-quality
X-ray data from Chandra.  We plot in Figure \ref{statistics.fig} the
distribution of the X-ray luminosity in the 0.5-2.0 keV energy band
versus redshift and highlight with red stars the five HIFLUGCS clusters
that are known to host a radio mini-halo;  we note that all
lie on top of the distribution.
For comparison, we also show other X-ray selected samples that cover
different $L_{\rm X}$-$z$ distributions: the observed Massive Cluster
Survey \citep[MACS,][]{macs_2010}, and the predicted eROSITA
thresholds \citep{erosita_2012}.

To derive the minimum X-ray luminosity of clusters with detectable
radio mini-halos in low-frequency surveys, we assume that mini-halos
have a typical spectral index $\alpha=1.2$
\citep[e.g.,][]{Giacintucci_2014a} and that consequently their 140~MHz
radio power has the same scaling with the X-ray luminosity as that at
1.4~GHz (Eq. \ref{Pradio-Lx.eq}).
The results for the different survey performances adopted in Table 1
are shown in Figure \ref{statistics.fig} (left panel) assuming a
typical magnetic field of B=1 $\mu$G, and the effect of varying
the magnetic field between B=1 $\mu$G and 30 $\mu$G are shown in
Figure \ref{statistics.fig} (right panel) for the SKA1-MID survey
alone.
Our main notes are listed below.\\
$\bullet \,$ The radio selection limits have different shapes than the X-ray flux-limited selections. This is important because
future radio surveys will allow us to reach much higher redshifts
than those of X-ray cluster selections with
similar sensitivities in the local universe.\\
$\bullet \,$  A consistent fraction of the clusters selected and
characterized by eROSITA will be observable by SKA and LOFAR in
various redshift ranges.  This is important because it will allow us
to build large cluster samples that can be exploited to study not only
the occurrence of radio mini-halos, but also the radio and X-ray
properties with an accuracy sufficient to study the interplay between
nonthermal and thermal cluster components.\\
$\bullet \,$ Radio surveys can be used to identify the
first SCC in the universe, that is, the highest redshift mini-halos with
associated X-ray emission (particularly with SKA1-LOW at $z>0.6$, see
Sect. 3.3). This may place constraints on the formation of SCC and
related radio emission.

\subsection{How many radio mini-halos await discovery?}

We estimated the maximum number of radio
mini-halo candidates that might be discovered in future surveys as a
function of redshift $z$ by integrating the radio luminosity function
(RLF) of mini-halos ${ d N_{\rm MH} }/(dP dV) $ over radio
luminosity and $z$,
\begin{equation}
N_{\rm MH}^{\Delta z} =
\int_{z_1}^{z_2} dz^{\prime}
\left( {{dV}\over{dz^{\prime}}} \right)
\int_{P_{\rm min}(z^{\prime})}
dP {{d N_{\rm MH} }\over{dP dV}}
\label{nmh.eq}
,\end{equation}
\noindent
where the minimum radio luminosity detectable at a given $z$, $P_{\rm min}$, can be
estimated from Eq. \ref{fmin.eq}.  

Our basic assumptions to calculate the RLF in Eq. \ref{nmh.eq} are
that
\\
(1) every SCC
cluster \citep[as defined by][]{Hudson_2010} hosts a radio mini-halo, 
and this mini-halo follows the observed $P_{\rm radio}$-$L_{\rm X}$
correlation (Eq. \ref{Pradio-Lx.eq}),
and\\
(2) the fraction of SCC clusters does not evolve strongly with
redshift.

These are two strong assumptions. The first assumption is supported by
\cite{Giacintucci_2017} at least for massive systems. These
authors indeed found that about 80\% of massive CC clusters host a
mini-halo. On the other hand, the occurrence of mini-halos in less
massive CC clusters is currently unconstrained.  The second
assumption, still debated in the literature
\citep[e.g.,][]{Santos_2010,Samuele_2011}, is motivated for CC systems
with a luminosity above a few $\times 10^{44}$erg s$^{-1}$, but
current data do not allow us to observe this correlation at lower X-ray
luminosities, that is, in clusters that contribute to the number
counting of mini-halos from Eq. \ref{nmh.eq} (see Figure
\ref{statistics.fig}).

Under these assumptions, the RLF of mini-halos per
sky area surveyed in steradians is \citep[see Eq. 3.2 of][]{Gitti_2015}
\begin{equation}
 {{d N_{\rm MH} }\over{dP dV}} = {\rm f}_{\rm SCC} \, {{d N_{\rm cl}
  }\over{dL_{\rm X} dV}} \, {{dL_{\rm X}} \over{dP_{\rm radio}}} 
\label{rlf.eq}
,\end{equation} 
\noindent
where ${d N_{\rm cl} }/(dL_{\rm X} dV) = \phi (L_{\rm X}, z)$ is
the X-ray luminosity function (XLF) of galaxy clusters, ${\rm f}_{\rm
  SCC}$ is the fraction of clusters with SCC \citep[we adopted
$\sim$0.4, estimated for the HIFLUGCS sample by][]{Hudson_2010} and
${dL_{\rm X}}/{dP_{\rm radio}}$ is obtained from Eq. \ref{norm.eq}.

We considered the evolving XLF derived from the
high-redshift X-ray--selected 160 Square Degree {\it ROSAT} Cluster
Survey (160SD) by \cite{Mullis_2004} in a $\Lambda$-dominated
universe. 
We assumed the local XLF determined from the REFLEX survey, which is
the only local sample for which the XLF in a $\Lambda$-cosmology has
been explicitly measured \citep[][]{Bohringer_2002}, and the evolution
estimated through the maximum likelihood analysis using the 66
clusters at $z>0.3$ in the 160SD sample, adopted by
\citet{Mullis_2004} itself as the less biased results
\citep[for more details on the derivation of the analytic XLF, we
refer to][]{Mullis_2004}.

In particular, we adopted \citep[see Eq. 5 of][with $\alpha=1.69$
taken from their Table 1]{Mullis_2004} 
\begin{equation} 
\phi (L_{\rm X}, z) = \phi^*(z) \left(\frac{L_X}{L_X^*(z)} \right)^{-1.69}
\exp \left(-\frac{L_X}{L_X^*(z)} \right) \frac{1}{L_X^*(z)}
,\end{equation}
where 
$
\phi^*(z) = 2.9 \times 10^{-7} \left( (1+z)/(1+z_0) \right)^{0.6} \, {\rm Mpc}^{-3}
$, and
$
L_X^*(z) = 2.6 \times 10^{44} \left( (1+z)/(1+z_0) \right)^{-2.1} \, {\rm erg \, s}^{-1}
$,
$z_0$ being the median redshift sampled by the local XLF ($z_0$=0.08
for the REFLEX survey).

%%%%

Operatively, for each redshift interval $(z_1,z_2),$ we numerically solved the integral in the form

\begin{equation}
N_{\rm MH}^{\Delta z} =
\int_{z_1}^{z_2} dz^{\prime}
\left( {{dV}\over{dz^{\prime}}} \right)
\int_{P_{\rm min}(z^{\prime})}
dP_{\rm radio} \, {\rm f}_{\rm
  SCC} \, \phi (L_{\rm X}, z^{\prime}) \, \frac{1}{a} \, \frac{L_X}{P_{\rm radio}}
,\end{equation}
\noindent
where the ratio $L_X/P_{\rm radio}$ is calculated from Eq. \ref{norm.eq}, and
$a$ is the best-fit parameter of the correlation in
Eq. \ref{Pradio-Lx.eq}. 

The uncertainty on $N_{\rm MH}^{\Delta z}$ is mainly driven by the
$P_{\rm radio}$-$L_{\rm X}$ correlation. The internal accuracy of the
XLF measurements is approximately $\pm 10\% - 20\%$ (estimated from
the $\pm 1 \sigma$ excursion of the error envelopes in Fig. 4 of
Mullis et al. 2004), and the systematics are also small. For example,
the results of the BCS survey \citep{Ebeling_1997} vary by a maximum
of about $\pm 25 \%$ relative to the Schechter fit of the REFLEX
assumed in this work \citep{Mullis_2004}. Therefore, to estimate the
uncertainty on $N_{\rm MH}^{\Delta z}$ we fixed the cluster XLF and
performed Monte Carlo simulations by considering 2000 random choices
of the parameters $a$ and $b$, and report the 1$\sigma$ error.

\begin{figure*}
\centerline{
\includegraphics[bb=18 144 592 718, scale=0.46]{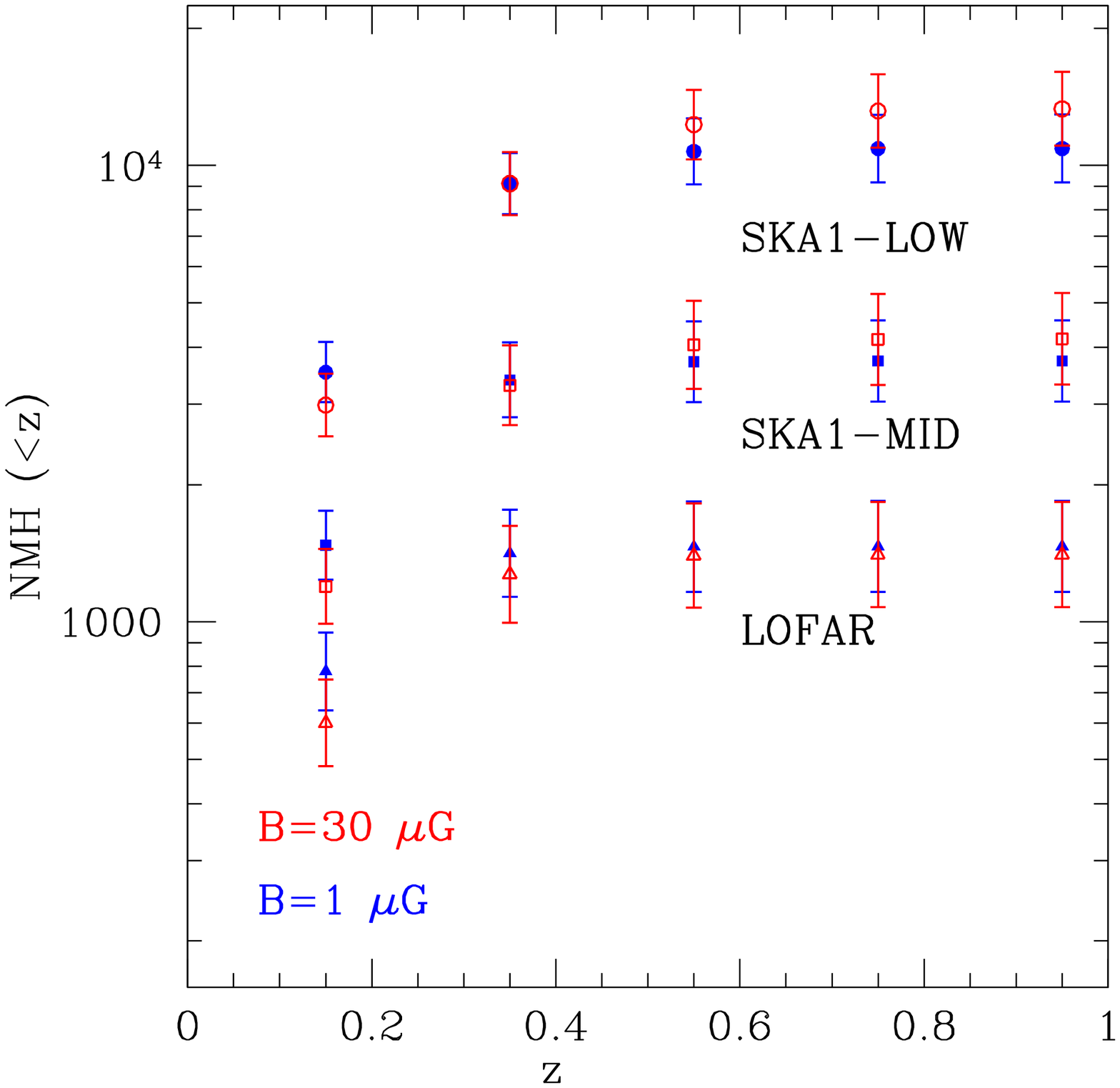}
\includegraphics[bb=18 144 592 718, scale=0.46]{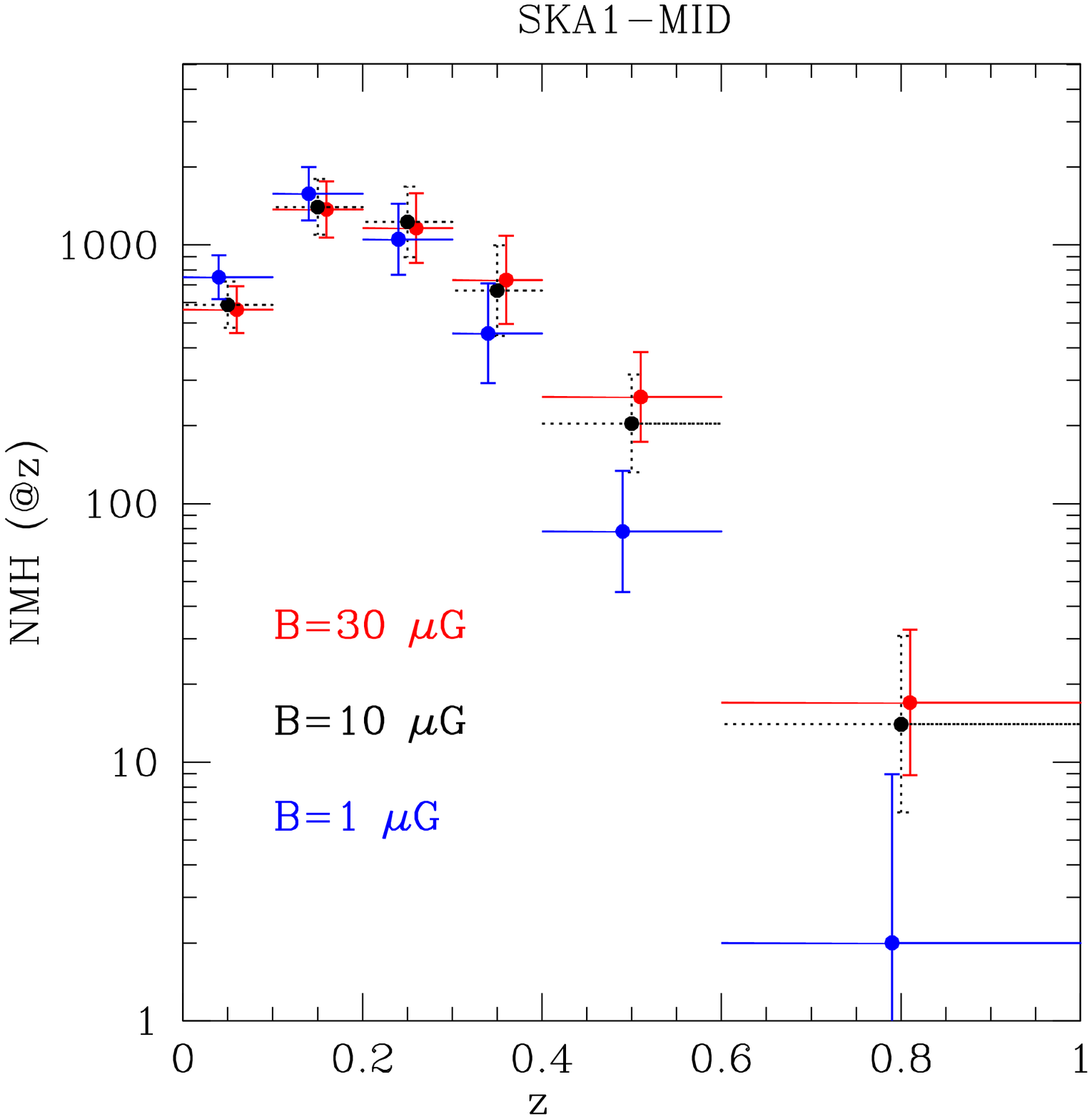}
}
\centerline{
\includegraphics[bb=18 144 592 718, scale=0.46]{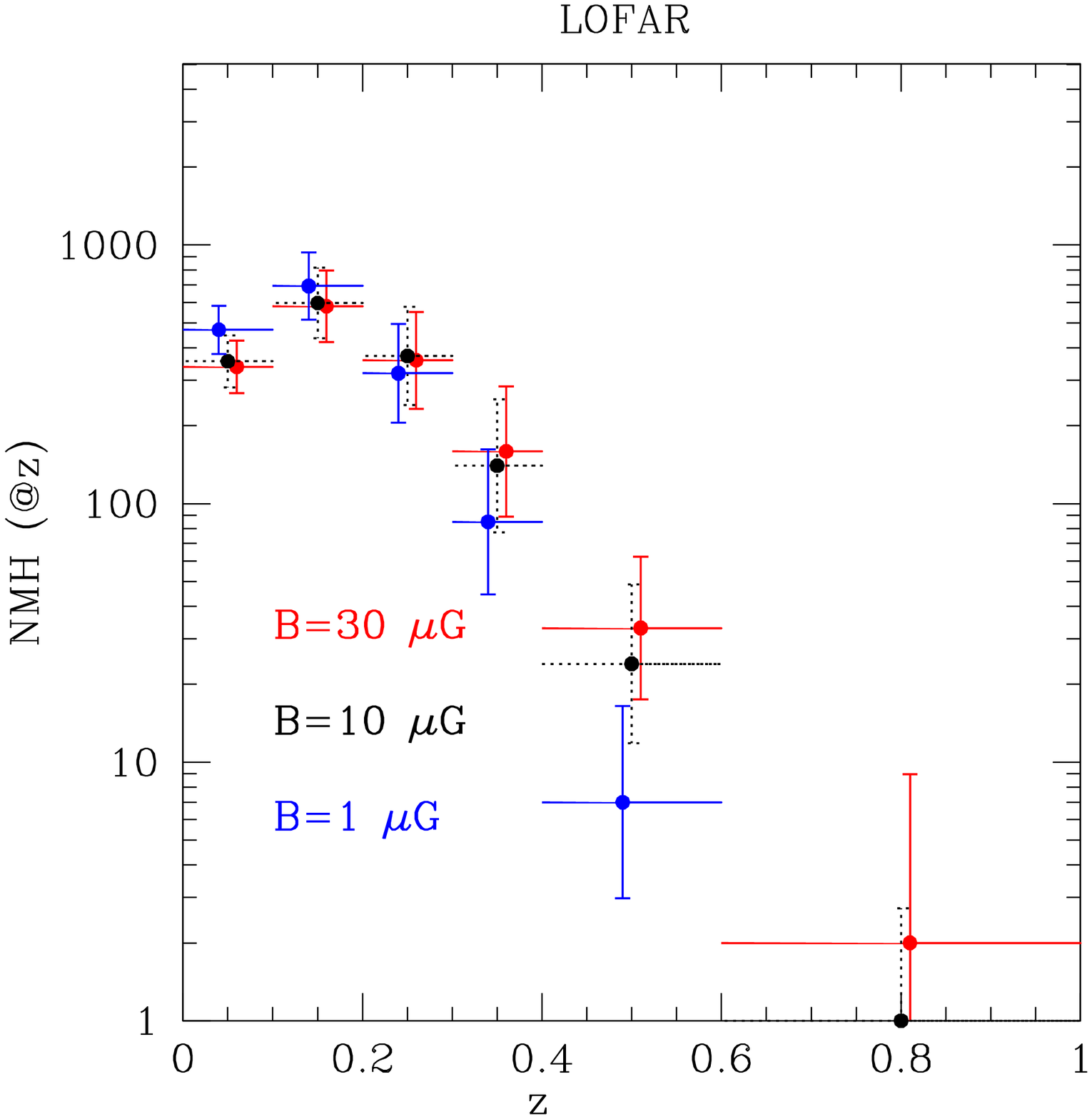}
\includegraphics[bb=18 144 592 718, scale=0.46]{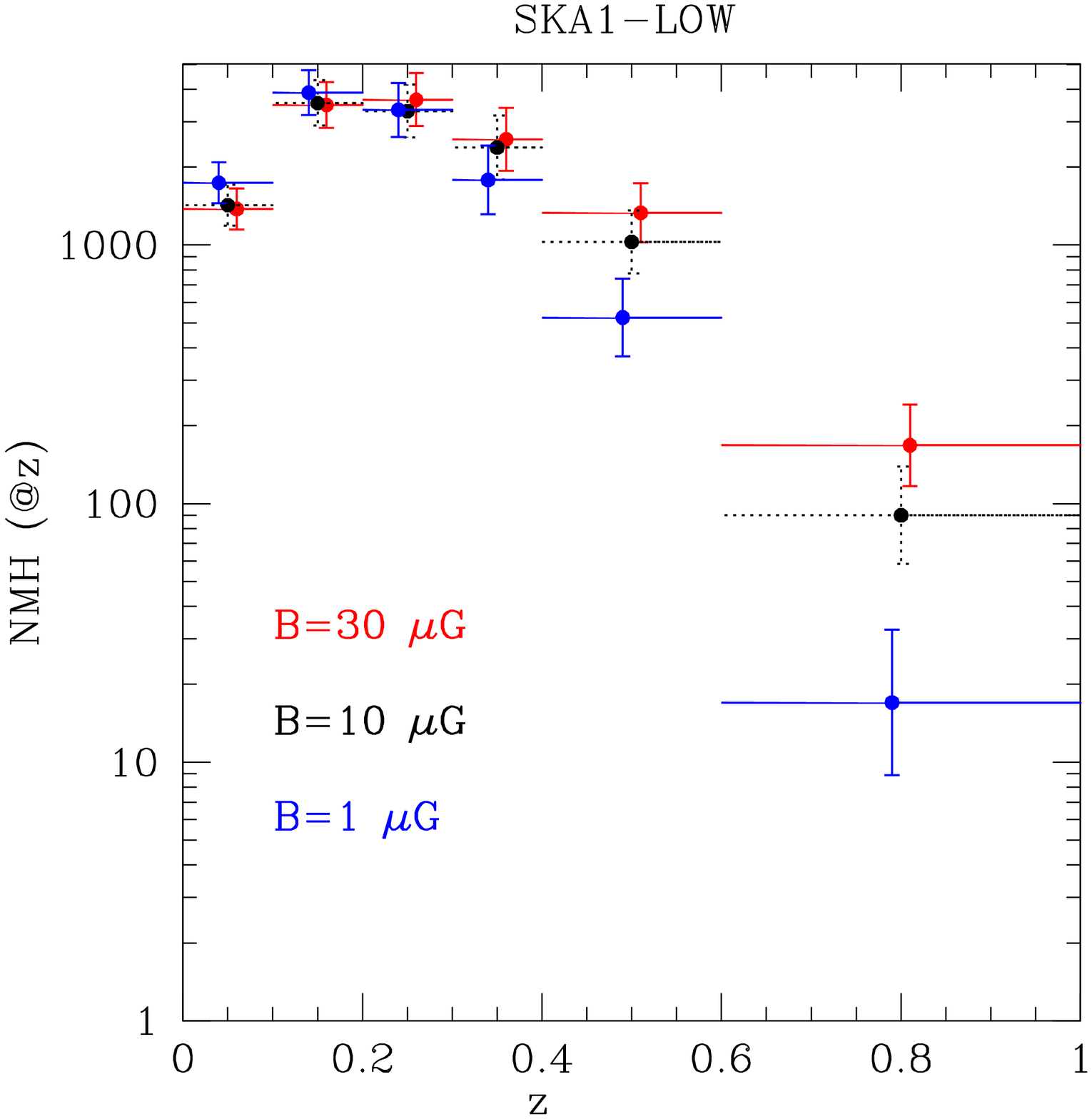}
}
\caption{\label{NMH.fig} {\it Top left panel:} Integral number of
  radio mini-halo candidates ($3 \pi$ sr) observable at 140~MHz by
  LOFAR (triangles) and SKA1-LOW (circles), and at 1.4~GHz by SKA1-MID
  (squares) as a function of redshift, estimated by assuming two
  reference values of $B=1\, \mu$G (blue) and $B=30\, \mu$G (red).
  The error bars indicate the 1$\sigma$ uncertainty driven by the
  $P_{\rm radio}$-$L_{\rm X}$ correlation (Eq. \ref{norm.eq}).  {\it
    Top right panel:} Differential number of radio mini-halo
  candidates ($3 \pi$ sr) observable at 1.4~GHz by SKA1-MID in
  different redshift bins, estimated by assuming three reference
  values of $B=1\, \mu$G (blue), $B=10\, \mu$G (black), and
  $B=30\, \mu$G (red). The vertical error bars indicate the 1$\sigma$
  uncertainty driven by the $P_{\rm radio}$-$L_{\rm X}$ correlation
  (Eq. \ref{norm.eq}), and the horizontal bars indicate the size
  of the redshift bin. The points corresponding to the three different
  $B$-field values in the same redshift bin are slightly shifted on
  the $x$-axis to improve clarity.  {\it Bottom left panel:} Same as
  top right panel, but showing the results for LOFAR at 140~MHz.  {\it
    Bottom right panel:} Same as top right panel, but showing the
  results for SKA1-LOW at 140~MHz.  }
\end{figure*}

We calculated the {\it \textup{integral}} number of expected radio mini-halo
candidates as a function of redshift by assuming $B$ between 1 and 30
$\mu$G.  Our predictions at 1.4~GHz (with SKA1-MID) and 140~MHz (with
LOFAR and SKA1-LOW) are shown in Figure \ref{NMH.fig} (top left
panel).
In particular, we estimated that all-sky surveys with SKA1 may be able
to detect up to $\sim$4000 and $\sim$10$^4$ new mini-halo candidates
out to redshift $z \sim$1 at -MID and -LOW frequency, respectively.
While we wait for SKA, its pathfinder LOFAR may already have the
capabilities to discover up to $\sim$1400 new mini-halo candidates,
which represents an enormous improvement with respect to the current
statistics.

These numbers refer to the assumption that all CC clusters host a
mini-halo, regardless of the mass and dynamical status of these
clusters.  In the context of hadronic models, this is probably
reasonable at some level. On the other hand, in the case of
reacceleration or leptonic models, the possibility of maintaining
radio-emitting electrons in the CC region depends on the level of
turbulence and on the fraction of the turbulent energy flux that is
damped into particle reacceleration.  If this turbulence is generated
by perturbations in the core that are induced by dynamics on larger
scales \citep[e.g.,][]{ZuHone_2013}, we may expect that the turbulent
level declines in less massive and less dynamically disturbed
systems. This will induce a declining fraction of mini-halos in CC, and
consequently, the number density of mini-halos would be significantly
smaller than that expected from our simple calculations. Although this
adds substantial uncertainties in the model expectations, it is also
true that a comparison between our simple expectations and future
surveys is a unique way to readily pinpoint the role played by
turbulence and reacceleration mechanisms in mini-halos, and raises the
possibility of distinguishing between different origins of these radio
sources.

In Figure \ref{NMH.fig} (top right, bottom left, and bottom right
panels) we report the \textup{\textup{}}differential number of mini-halos expected
to be observable in different redshift bins at 1.4~GHz (by SKA1-MID),
and at 140~MHz (by LOFAR and SKA1-LOW), respectively, estimated by
assuming three reference values of $B=1\, \mu$G, $B=10\, \mu$G, and
$B=30\, \mu$G.

Model expectations significantly differ at redshift higher than
0.6. The reason is that in the case of weak magnetic fields an
increasing fraction of the nonthermal luminosity is channelled into
inverse Compton emission at higher redshifts, making mini-halos
progressively less luminous and thus more difficult to observe. Based
on our calculations, the future SKA1 radio surveys will allow us to
distinguish among different assumptions on the magnetic field
intensities in the mini-halo region.
For example, an SKA1-MID detection of more than 10 mini-halos at
$z>0.6$ will suggest a stronger magnetic field ($B \geq
10$-$30 \, \mu$G), but an SKA1-LOW non-detection of mini-halos
above the same redshift will indicate a lower value of
$B < 1 \, \mu$G.
On the other hand, LOFAR observations are already able to provide some
constraints. For example, a LOFAR detection of more than 20 mini-halos
in the redshift bin $z$=0.4-0.6 will suggest $B \geq 30 \mu$G. In any
case, we expect fewer than 10 mini-halo candidates detected by LOFAR at
$z>0.6$, regardless of the value of the magnetic field.

This argument on the magnetic field is straightforward in the case of a
hadronic origin of mini-halos. On the other hand, in reacceleration
models the increase of inverse Compton losses with redshift makes
reacceleration more difficult. The decline in the
expected number of mini-halos in the case of weak fields may therefore be much
stronger than predicted in Figure~\ref{NMH.fig}.
All these caveats require future theoretical investigations
to better exploit future radio surveys.

\section{Discussion and conclusions}

We started to consider the capabilities of future surveys
in detecting mini-halos by considering facilities operating at low
(LOFAR, SKA-LOW) and higher (SKA-MID) frequencies.  We obtained
expectations based on the observed correlation between the radio and
X-ray luminosity measured in mini-halo clusters and on its
extrapolation toward systems with lower X-ray luminosities and higher
redshift.  We showed here that future surveys with the SKA and its
pathfinder LOFAR have the potential of detecting a large number
($\sim 10^3 - 10^4$) of radio mini-halo candidates up to redshift
$z \sim 1$.
This expectation is optimistic because it is based on the assumption
that every CC cluster hosts a mini-halo, an assumption that is
currently supported only for massive systems at redshift $\le$ 0.35
\citep{Giacintucci_2017}. On the other hand, if mini-halos are
generated by turbulent reacceleration, we might expect a significant
drop of their occurrence in less massive CC and at higher redshift.
The fraction of nonthermal emission that is channelled into
synchrotron radiation declines with redshift in a way that depends on
the magnetic field in the mini-halo region. We have shown that this
effect generates a decline in luminosity function of mini-halos
with redshift and that future surveys with the SKA consequently have
the potential of constraining the magnetic field in these systems.

As a caveat for these estimates, we stress that the contamination
of point sources, particularly of the central BCG, and the dynamic
range of the radio images are important drawbacks for the detection of
the diffuse emission of mini-halos at high redshift.  However, the SKA is
planned to achieve a high dynamic range and to be able to reach
angular resolutions (a few arcseconds) that will allow us to accurately subtract the point
sources by directly removing their visibilities in the {\it
  uv}-plane. Therefore, the residual flux, if present, is
assumed to be entirely due to the diffuse emission. The angular
size of the mini-halo relative to the resolution of the instrument is
therefore not expected to be an issue for the identification of mini-halos.

These caveats are more important at lower frequencies. First, the
LOFAR survey LoTSS is being carried out at a nominal resolution of 6
arcsec \citep{Shimwell_2017}. This resolution may not
be sufficient to accurately subtract discrete sources at higher
redshift. Nevertheless, it is also true that higher-resolution images
can be generated from the LoTTS data with an appropriate use of the
weighting and tapering functions applied to the visibilities during
the reduction process, thus controlling the beam shape. Furthermore,
the mini-halo candidates selected by the LoTTS could be followed-up at
higher resolutions with dedicated, pointed LOFAR
observations. Cross-checks with X-ray catalogs will also help to
identify AGNs.
Second, at low radio frequencies, old bubbles from the central AGNs
become more prominent and may provide a significant contamination to
the truly diffuse radio emission from the mini-halos. However, these
observational issues and the details of the data analysis are beyond
the aim of this paper.

We also discussed the synergy of radio surveys with future X-ray
missions, showing that a consistent fraction of the clusters selected
and characterized by eROSITA will be observable by SKA and LOFAR in
various redshift ranges. This will allow us to build large statistical
samples of galaxy clusters that can be exploited to study not only the
occurrence of radio mini-halos, but also the radio and X-ray
properties with an accuracy sufficient to investigate the interplay
between nonthermal and thermal cluster components.
In particular, according to numerical simulations \citep{ZuHone_2013},
turbulence with $\delta v \sim$ 50-200 km s$^{-1}$ on spatial scales
below $\sim$50-100 kpc in the CC region might be sufficient to
reaccelerate seed relativistic electrons producing radio mini-halos.
Such turbulence levels are easily detectable with typical exposures of
the Athena X-ray Integral Field Unit
\citep[X-IFU,][]{Barcons-Athena_2017}, which is expected to resolve
gas velocities of tens of km s$^{-1}$ in the cluster cores on 5 arcsec
scales \citep{Ettori-Athena_2013}, corresponding to $\sim$40 kpc at
z$\sim$1.
The synergies of the radio surveys discussed in this work with future
Athena X-IFU observations and theoretical studies is essential in
establishing physical nature of radio mini-halos.

\section*{Acknowledgments}

We would like to thank the anonymous referee for providing
constructive comments and raising interesting points that improved the
presentation of our work.  We acknowledge support from contract
ASI-INAF 2017-14-H.0.  MG, GB and RC acknowledge support from
PRIN--INAF2014.  SE acknowledges the financial contribution from
contracts ASI-INAF I/037/12/0, ASI 2015-046-R.0.

\bibliographystyle{aa}
\bibliography{32749corr_final}

\end{document}